\documentclass[sigconf,nonacm]{acmart}

\usepackage{geometry}% http://ctan.org/pkg/geometry
\usepackage{graphicx}% http://ctan.org/pkg/graphicx
\usepackage{paralist}
\usepackage{xspace}
\usepackage{algorithm}
\usepackage{algpseudocode}
\usepackage{multirow}
\usepackage[table]{xcolor}
\definecolor{paleblue}{RGB}{233,242,251}
\usepackage{url}
\usepackage{balance}

\usepackage{breqn}
\usepackage{booktabs}
\usepackage[most]{tcolorbox}
\usepackage{listings}

\lstdefinestyle{promptstyle}{
    basicstyle=\ttfamily\footnotesize,
    breaklines=true,
    breakatwhitespace=false,
    breakindent=0pt,
    breakautoindent=false,
    columns=fullflexible,
    keepspaces=true,
    showstringspaces=false,
    upquote=true,
    tabsize=2
}

\newtcolorbox{promptsectionbox}{
    enhanced,
    breakable,
    colback=white,
    colframe=black,
    boxrule=0.7pt,
    arc=0pt,
    outer arc=0pt,
    left=7pt,
    right=7pt,
    top=8pt,
    bottom=8pt,
    before skip=8pt,
    after skip=14pt
}

\AtBeginDocument{%
  }

\begin{document}

\title{EgoVeil: Personalized Privacy Reasoning for Egocentric Video}
\author{Xue Wang}
\orcid{0000-0002-4551-4932}
\affiliation{%
  \institution{University of California, Los Angeles}
  \city{Los Angeles}
  \state{CA}
  \country{USA}
}
\email{xw526@ucla.edu}

\author{Yang Zhang}
\orcid{0000-0003-2472-6968}
\affiliation{%
  \institution{University of California, Los Angeles}
  \city{Los Angeles}
  \state{CA}
  \country{USA}
}
\email{yangzhang@ucla.edu}

\def\systemname {\textit{EgoVeil}\xspace}

\begin{teaserfigure}
  \includegraphics[width=1\textwidth]{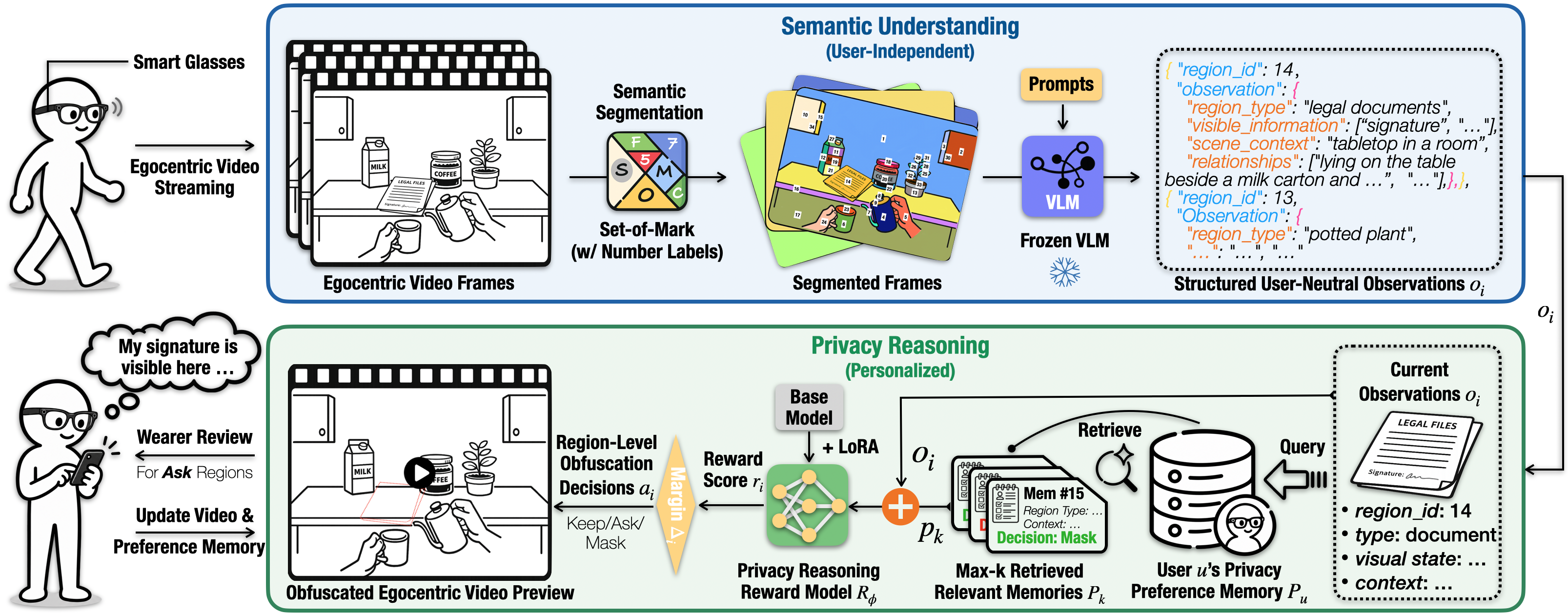}
  \caption{Our personalized privacy preservation system for egocentric video. (Top) The system first performs user-independent semantic understanding by segmenting egocentric video frames and using a frozen vision-language model to produce structured, user-neutral observations. (Bottom) Personalized privacy reasoning combines each current observation with relevant prior preferences retrieved from the wearer’s privacy preference memory to make region-level Keep, Ask, or Mask decisions. Ask regions are presented to the wearer for review, and the resulting feedback updates the obfuscated video and preference memory for future decisions.}
  \label{fig:teaser}
\end{teaserfigure}

\begin{abstract}

Recent advances in vision-language models (VLMs) enable AI systems to interpret egocentric video. Yet continuous visual capture raises privacy concerns because sensitive content is pervasive and varies across users and contexts. We present \systemname, a personalized privacy reasoning system that determines which visual content to obscure based on individual preferences while retaining context useful for downstream assistance. It converts localized visual regions into structured semantic observations and combines a privacy reasoner with an explicit user-specific memory for personalized privacy reasoning. This design supports personalization and continual adaptation as preference evidence accumulates. We evaluate on in-the-wild data from 10 smart-glasses users under a cross-user chronological protocol. \systemname achieves 91.93\% balanced accuracy and 84.42\% Mask recall, outperforming direct VLM baselines and other controlled privacy reasoning baselines. Our results show that personalized, adaptive privacy reasoning better aligns privacy preservation with individual preferences while retaining the visual context needed for effective smart-glasses assistance.

\end{abstract}




\maketitle

\section{Introduction}
Smart glasses offer a promising interface for vision-based AI assistance in everyday use~\cite{zhang2025through, yang2025egolife}. Their cameras capture egocentric video from the user’s perspective, providing rich contextual information about the surrounding environment and ongoing activities. Recent advances in vision-language models (VLMs) have further improved AI systems’ ability to understand and reason about egocentric video, enabling a broad range of assistive applications~\cite{luo2026openmmego, suglia2024alanavlm, xiao2026egoblind}. However, the rich visual context captured in egocentric video may also contain information that users consider private~\cite{bhardwaj2024focus, li2025egoprivacy}. This creates a fundamental tension between utility and personalized privacy preservation~\cite{alharbi2019mask, wang2025invisibility, orekondy2018connecting}, as downstream AI systems benefit from access to rich visual context for analysis, while decisions about what information to retain for these analyses depend on each user's privacy preferences.

Prior research on privacy preservation has investigated various mechanisms for wearable cameras across the imaging process. Many existing approaches rely on predefined visual categories or contexts to guide capture control or selective obfuscation~\cite{raina2023egoblur, li2025extra, steil2019privaceye}. While effective for addressing known privacy risks, this framing does not fully account for how privacy preferences vary across users and situations. Multiple prior studies have shown that visual privacy preferences are highly individualized and context-dependent~\cite{orekondy2017towards, hoyle2020privacy, banerjee2026personalized}. This limitation makes fixed privacy categories insufficient for open-ended egocentric settings and calls for privacy preservation mechanisms that reason about visual information in relation to both its context and an individual user’s preferences.

To better understand how people make personalized privacy decisions in their lives, we conducted an in-the-wild study with 10 participants, where each participant wore smart glasses and recorded egocentric video clips during their daily activities. Participants were then asked to review their own recordings, identify privacy-sensitive visual regions, and explain the reasoning behind their decisions. We found that privacy decisions were rarely determined by object categories alone. Instead, their privacy decisions often depended on what a visual region revealed in context. Visually similar content could therefore receive different privacy decisions across users and situations. We further observed that users did not always articulate all relevant preferences in advance, but progressively refined and expanded them as they encountered new situations. 

In this work, we introduce \systemname, a personalized privacy reasoning system for egocentric video. It separates user-independent visual understanding from personalized privacy reasoning. We first convert the localized visual regions into structured semantic observations using a locally deployed VLM without access to user preferences. These observations are then passed to a locally deployed privacy reasoning model, adapted via LoRA fine-tuning, which provides task-specific privacy reasoning. To personalize the privacy reasoning for each region, \systemname also maintains an explicit user-specific preference memory and retrieves prior confirmed decisions pertinent to the current observation. These two components work together, the privacy reasoning model provides a learned basis for privacy decisions given the observations and relevant memories, while the retrieved memory supplies user-specific evidence reflecting the current user's expressed preferences. As the user confirms or corrects more privacy decisions made by the system, these preferences are added to the memory and can inform subsequent decisions without retraining the model.

To understand the personalized privacy requirements and evaluate the system performance, we collected an in-the-wild egocentric video dataset from 10 smart-glasses users across diverse everyday activities and environments. Participants annotated privacy-sensitive visual regions in their own recordings and provided rationales explaining their decisions, yielding over 12000 region-level privacy annotations and 100 utility-related questions per user on average. We evaluate \systemname under a chronological cross-user protocol, where each user is held out during training of the privacy reasoner and begins with an empty preference memory. Only preferences obtained from earlier clips are available when later clips are processed, simulating deployment to a previously unseen user and subsequent personalization over time. Our proposed system achieves 91.93\% balanced accuracy and 84.42\% Mask recall, outperforming both direct VLM baselines and controlled reasoning ablations. We further find that performance improves as more user-specific preference history becomes available in the controlled comparison evaluations, with balanced accuracy increasing from 56.57\% without preference memory to 95.75\% with the full preceding history. Finally, we show that the \textit{Ask} mechanism can effectively target decisions that benefit from user feedback, outperforming random selection and enabling more efficient confirmation of user-specific preference evidence.

Below we list our key contributions:
\begin{itemize}
    \item First characterize personalized visual privacy decisions in everyday egocentric video through an in-the-wild study, yielding a dataset of region-level privacy decisions, rationales, and utility questions.
    
    \item Develop \systemname, a personalized privacy reasoning system that aligns region-level decisions with user-specific preferences through shared privacy reasoning and explicit preference memory.
    
    \item Evaluate the system under a chronological cross-user protocol, demonstrating the complementary roles of learned privacy reasoning and user-specific memory.
\end{itemize}

\section{Related Work}

\subsection{Visual Privacy Preservation through Obfuscation}
Visual privacy can be preserved before capture or after visual data have been recorded. Capture-time systems block or disrupt imaging through mechanisms such as mechanical shutters or illumination modulation~\cite{steil2019privaceye,zhu2017automating}. These approaches reduce privacy leakage at the source but also remove information that may support later sharing or analysis. Post-capture methods instead apply blurring, pixelation, masking, or learned visual degradation~\cite{wu2018towards,alharbi2019mask,li2017blur,wang2021infoscrub} to preserve some context while reducing privacy exposure. Early methods often transformed entire images or videos~\cite{boyle2000effects,neustaedter2006blur}, creating a trade-off in which weak transformations may leave sensitive cues visible, whereas stronger transformations reduce visual utility.

Selective obfuscation modifies only identified sensitive regions or attributes. Orekondy et al.~\cite{orekondy2018connecting} detect and selectively obscure privacy-sensitive regions; PrivObfNet~\cite{tay2024privobfnet} localizes regions associated with predefined privacy attributes; and Ilic et al.~\cite{ilic2024selective} obscure user-specified attributes while preserving temporal consistency for action recognition. These methods generally define sensitivity through predefined categories or attributes, which do not fully capture differences across users and contexts~\cite{wang2025invisibility}. Imago Obscura~\cite{monteiro2025imago} and Invisibility Cloak~\cite{wang2025invisibility} demonstrate the value of personalized obfuscation, but rely primarily on preferences or settings explicitly provided by users. \systemname instead combines contextual region descriptions with user-specific preference evidence accumulated through prior decisions and feedback.

\subsection{Privacy Management for Wearable Cameras}
Wearable cameras pose distinct privacy challenges because their first-person perspective and unobtrusive capture can record users, bystanders, and sensitive environments without explicit consent~\cite{dimiccoli2018mitigating,lee2020pedestrian,motti2015users,denning2014situ,hoyle2014privacy}. Early systems controlled capture by recognizing sensitive places or situations. PlaceAvoider~\cite{templeman2014placeavoider} learns user-specified sensitive locations, and PrivacEye~\cite{steil2019privaceye} combines egocentric scene and eye-movement cues to close a camera shutter when needed. Other systems obscure backgrounds, bystanders, faces, license plates, or other predefined types of identifiable information~\cite{alharbi2019mask,raina2023egoblur,li2025extra,khawaja2026see}. These approaches address common and anticipated risks but still define sensitivity primarily through fixed object classes.

As VLMs become integrated into smart glasses, wearable systems can infer high-level contextual information and answer questions about captured scenes~\cite{zhang2025through,li2025egoprivacy}. This broadens privacy risks beyond directly identifiable objects and makes privacy decisions more dependent on context. Users and bystanders may also hold different expectations about when visual information should be captured and shared~\cite{wang2026mind}. Existing systems use VLMs to convert egocentric images into textual descriptions~\cite{qiu2023egocentric,hiraoka2026edgevlm} or to inspect and edit sensitive content~\cite{dinh2026unsafe2safe}, but generally follow fixed privacy policies or manually specified rules. \systemname instead uses a VLM to describe what each localized region reveals in context and then applies a separate personalized reasoner informed by the current user's preference history.

\subsection{LLM-powered Personalization}
LLMs are generally trained to align with human preferences~\cite{ouyang2022training,christiano2017deep,bai2022training,rafailov2023direct}. Personalization research has therefore explored methods that tailor model behavior to individual users~\cite{nam2025learning,li2024personalized,wu2025aligning}. One common approach keeps user information explicit in profiles or histories. LaMP~\cite{salemi2024lamp} retrieves relevant entries from a user's history, and Zhang et al.~\cite{zhang2024guided} summarize personal context into concise preference descriptions. VisualLens~\cite{zhu2026visuallens} and PLVM~\cite{pham2025plvm} extend retrieval-based personalization to multimodal histories and personalized visual concepts. These approaches avoid per-user model training and keep preference evidence easy to update, but retrieval relevance may not correspond to personalization value. User histories may also be sparse or contain irrelevant, redundant, outdated, or conflicting records~\cite{du2026optimizing,zhang2025proper,chao2026stale,tao2026memconflict}.

Another approach encodes preferences into latent representations or model parameters. Poddar et al.~\cite{poddar2024personalizing} infer user-specific latent variables from a small number of preference annotations, and PReF~\cite{shenfeld2025language} represents individual preferences as combinations of shared reward functions. In multimodal settings, Pi et al.~\cite{pi2025personalized} fine-tune VLMs to recognize specific individuals and support personalized dialogue. These methods can capture patterns shared across users and adapt from limited feedback, but revising encoded preferences may require additional training. Other systems update user profiles or memories across interactions, including Accessibility Scout~\cite{huang2025accessibility} and PersonaVLM~\cite{nie2026personavlm}. Privacy-oriented systems such as Rescriber~\cite{zhou2025rescriber} and Think Twice~\cite{pias2026think} use LLMs to help users manage information disclosure.

Personalized alignment has been studied mainly in text-centered applications such as dialogue, recommendation, and question answering. Visual privacy additionally requires grounding user preferences in fine-grained visual regions and interpreting those regions within the current context. \systemname combines shared LoRA adaptation for privacy-task reasoning with explicit user-specific preference memory, allowing cross-user reasoning patterns and individual preference evidence to remain separate. Newly confirmed preferences can then inform subsequent decisions through memory without training separate model parameters for each user.

\section{Data Collection and Formative Observations}
\label{sec:user study}

We conducted an in-the-wild study to examine how participants make privacy decisions and to collect data for system development and evaluation. Participants were asked to wear smart glasses and record activities from their daily routines. During the data annotation process, they reviewed their own recordings, identified visual regions they wanted obscured, and explained the reasoning behind each decision. Participants also confirmed question-answer pairs and moment descriptions about the recorded content to support the evaluation of downstream utility. The privacy annotations, rationales, and interview data informed our formative observations and the design of \systemname. The region-level privacy decisions and utility questions were used for model training and evaluation subsequently.

\subsection{In-the-Wild Egocentric Data Collection}
\label{sec:dataset}

\subsubsection{Participants and Study Procedure}
We recruited 10 participants (8 female) to record egocentric videos during their daily activities using Ray-Ban Meta smart glasses~\cite{raybanstories2021}. We did not assign predefined tasks. Participants were free to record moments when they would naturally use the smart glasses. They were encouraged to capture a diverse range of activities, environments, and situations while minimizing overlap across recordings to avoid collecting substantially similar content. Each participant was asked to record at least $30$ one-minute video clips at $30$ FPS and $1080$p resolution. 

\begin{figure*}[t]
    \centering
    \includegraphics[width=1\linewidth]{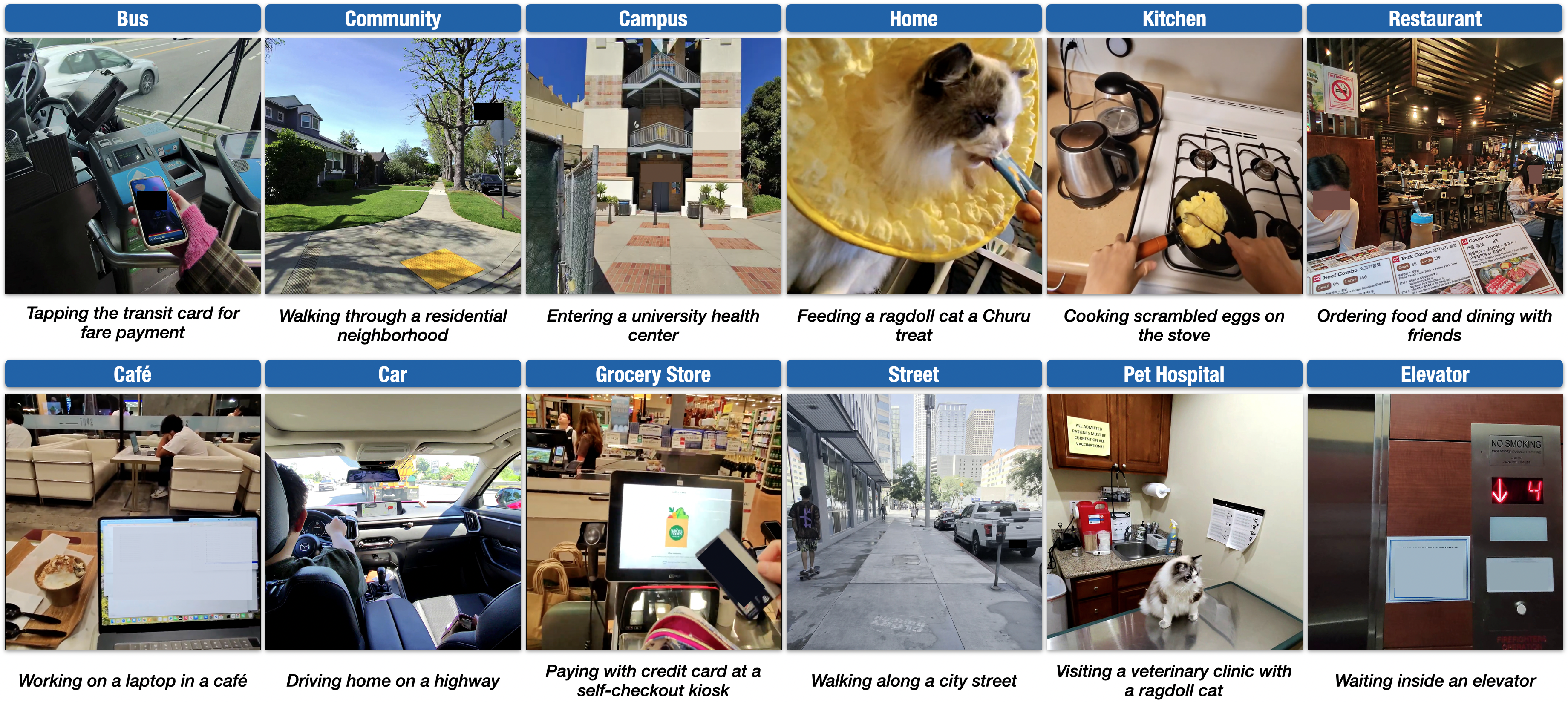}
    \caption{Examples from the collected in-the-wild egocentric dataset. The videos capture a diverse range of everyday environments and activities. Participants recorded these videos naturally using Meta Ray-Ban smart glasses during their daily lives. Regions obfuscated for anonymity.}
    \label{fig:dataset_example}
\end{figure*}

On average, each participant recorded 36.6 clips (SD = 7.68), totaling approximately 6.1 hours of egocentric video. The collected videos covered a wide range of real-world settings, such as public transportation, campuses, and restaurants, and activities like shopping, walking, and cooking. Figure~\ref{fig:dataset_example} shows some examples from the collected dataset.

\subsubsection{Annotation Preprocessing and Interface}
\begin{figure}[t]
    \centering
    \includegraphics[width=\linewidth]{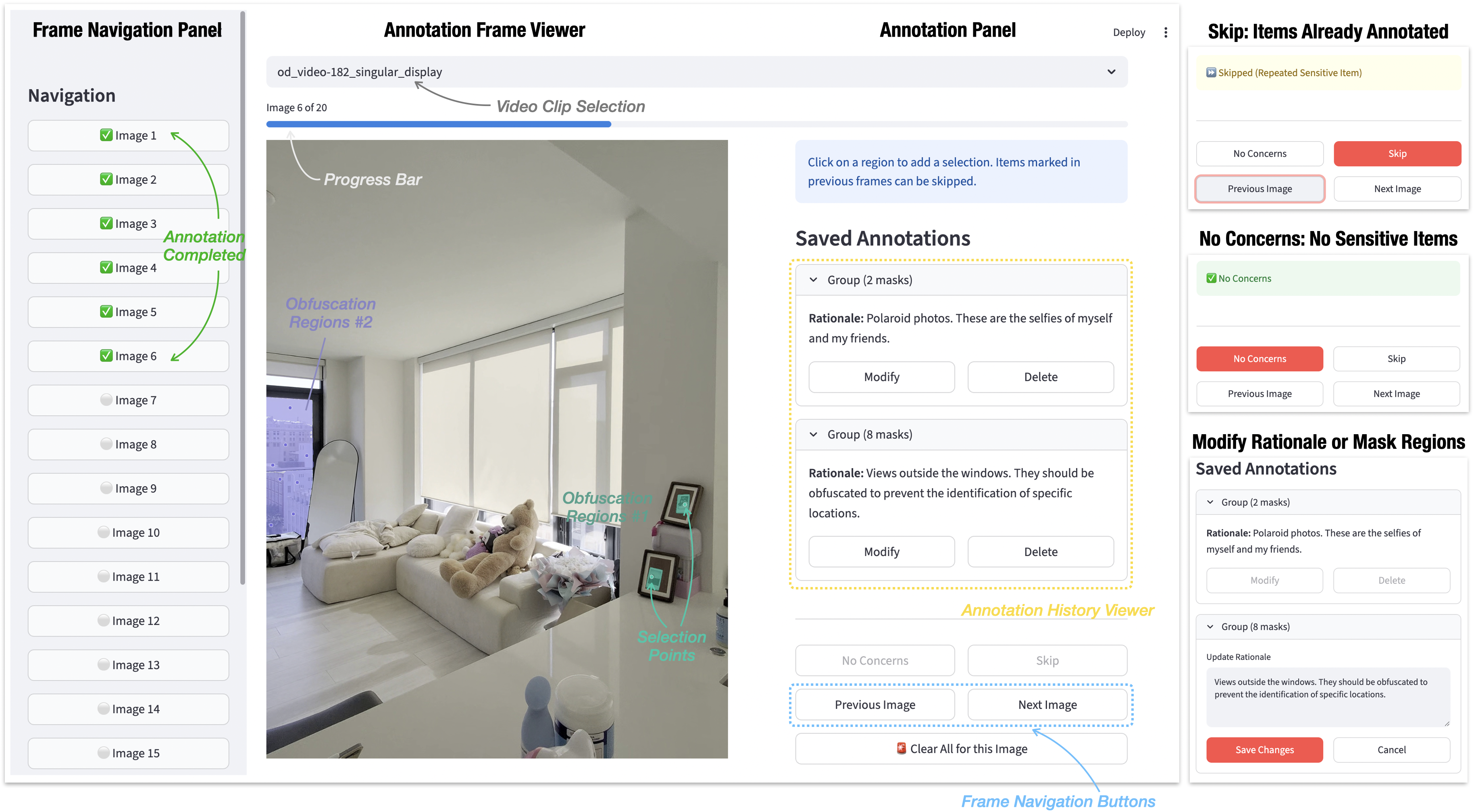}
    \caption{Privacy annotation interface. Participants navigate frames using the frame navigation panel (left).  For each keyframe, they can select one or more segmentation regions corresponding to a privacy-sensitive item, provide a textual rationale, and review previously saved annotations in the annotation panel (center). Clicking a selected region again can deselect it. The interface also supports marking frames with no privacy-sensitive items, skipping items already annotated in previous keyframes (top right), and modifying previously annotated mask regions and rationales (bottom right).}
    \label{fig:annotation_interface}
\end{figure}

After completing the data collection, we conducted individual annotation interviews in which participants reviewed and annotated their own recordings. There are two forms of annotations -- \textit{privacy annotations} identified the visual regions participants wanted to obscure and document the rationale behind each decision, while \textit{utility annotations} asked participants to prepare question-answer pairs and moment descriptions for each video clip, which were later used to evaluate whether privacy-preserved videos retained the information needed for downstream tasks.
To reduce annotation effort while supporting region-level decisions, we preprocessed the recorded clips and developed a locally hosted web application, shown in Figure~\ref{fig:annotation_interface} to guide participants through the annotation workflow.

\paragraph{Privacy Annotation} \label{para:privacy anno}
Each one-minute video clip contains approximately $1,800$ frames. Reviewing every frame would have imposed a substantial annotation burden because adjacent frames often contain highly similar visual content due to the high frame sampling rate. However, annotating an entire clip as a single unit could make it difficult for participants to precisely localize their privacy concerns and could cause them to overlook details during video playback. We therefore applied a keyframe selection procedure to all recorded clips (see Appendix~\ref{app:keyframe} for the processing details). The procedure examined every decoded frame to identify distinct and representative frames without imposing a fixed keyframe budget, allowing the number of selected frames to vary according to the visual richness of each clip.
Across all clips, the procedure selected an average of $56.24$ keyframes per one-minute clip (SD = $12.81$). Participants annotated the selected keyframes using our web interface. After completing the keyframe annotations, they reviewed each full video clip and identified any additional moments containing visual regions they wanted obscured. Corresponding frames were then added to the annotation set. This preprocessing procedure was used only to support data annotation and was not part of the \systemname inference pipeline.

For each selected frame, we generated category-agnostic visual regions using Semantic-SAM~\cite{li2024segment} with the pretrained Swin-L checkpoint at semantic level~$2$. We removed small and near-duplicate masks, suppressed redundant nested and strongly overlapping proposals. When the initial region set left substantial portions of the frame uncovered, we automatically prompted SAM~\cite{kirillov2023segment} with the ViT-H checkpoint in large uncovered areas to generate additional masks. The resulting regions were presented as spatial selection aids (see Figure~\ref{fig:semantic-understanding} middle for example), allowing participants to identify localized visual content without manually tracing region boundaries. Participants could select a single region or group multiple regions when they jointly represented the same privacy-sensitive item and shared one privacy decision and rationale. If the desired area was still not adequately covered by the available regions, participants could place a free point at that location, which was used as a prompt to SAM to generate an additional segmentation mask. Across all selected frames, the automatically generated regions covered an average of 93.03\% of the image area (SD=$5.39$) and provided 16.38 selectable regions per frame on average (SD=$9.36$). Most of the uncovered regions are due to hair occlusion or poor illumination during manual inspection.

We developed a locally hosted web application using Streamlit~\cite{streamlit} to facilitate the annotation process, as illustrated in Figure~\ref{fig:annotation_interface}. Participants could select multiple regions within a single annotation when the relevant content spanned several proposals or when the same privacy decision applied to multiple instances. If no region adequately covered the intended region, participants could mark the uncovered location with a free point, which was then used as a prompt to SAM to generate an additional segmentation mask. Participants were instructed to inspect each frame carefully and identify all visual content they considered sensitive. For each selected region or region groups, participants were required to provide a guided free-form rationale specifying the category of the item and explaining which visible content raised their privacy concern. If a frame contained no sensitive content, participants marked it as \textit{No Concerns}. Participants labeled a frame as \textit{Skip} when it contained no new privacy concern beyond content already annotated in an earlier frame from the same clip. This option reduced repeated annotations of recurring content and limited its overrepresentation in the dataset.

All annotations were saved and synchronized to local JSON files in real time, with a separate file maintained for each video clip. Each file contained the clip name and, for every annotated frame, the frame ID, frame label, selected regions, click coordinates, and corresponding rationales. A researcher was present throughout each session to answer procedural questions and request clarification when a selection or rationale was ambiguous. All study procedures were conducted under a protocol approved by our Institutional Review Board (IRB).

\paragraph{Utility Annotation}\label{para:utility anno}
Privacy annotations identify which privacy-sensitive content should be removed; equally important, however, is evaluating how such removal affects the informational richness of the remaining video for downstream utility tasks. To address this, we conducted a utility annotation study. Following the episodic memory annotation protocol from the Ego4D benchmark~\cite{grauman2022ego4d}, participants reviewed each video and confirmed two or three natural-language question-answer pairs for each video clip, following the Ego4D question templates to ensure consistent and systematic annotations across participants. 
Table~\ref{tab:utility_annotation} displays the question categories, representative examples, and their distribution in the dataset. 
Similar to the privacy annotation process, a researcher reviewed and confirm the answer for each question together with the participant. These question-answer pairs provide reference data for evaluating how privacy preservation affects the information available for downstream video understanding. We use these annotations to compare video question-answering accuracy between the original videos and their obfuscated counterparts.

\begin{table*}[t]
\centering
\begin{tabular}{lllc}
\toprule
\rowcolor{paleblue!90}
\textbf{Category} & \textbf{Example Questions} & \textbf{Answer} & \textbf{Proportion (\%)} \\
\midrule
\multirow{3}{*}{Object Localization}
& Where is my backpack? & Beside the chair. & \multirow{3}{*}{10.62} \\
& Where did I put my phone? & On the dining table. & \\
& In what location did I see the shopping cart? & Grocery store. & \\
\midrule
\multirow{5}{*}{Object Attributes}
& What did I put in the refrigerator? & A bottle of milk. & \multirow{5}{*}{47.39} \\
& How many drinks were on the table? & Three. & \\
& What food did I place on the plate? & Sandwiches. & \\
& What object is beside the laptop? & A notebook. & \\
& Was the laptop open or closed? & Open. & \\
\midrule
\multirow{3}{*}{Person Interaction}
& Who did I interact with while ordering food? & The cashier. & \multirow{3}{*}{5.02} \\
& Who did I talk to in the café? & My friend. & \\
& When did I interact with the cashier? & Before paying. & \\
\midrule
Place Identification
& Where was I when I saw the menu? & At the restaurant. & 7.92 \\
\midrule
\multirow{2}{*}{Temporal Reasoning}
& Where was my phone before I opened my laptop? & On the table. & \multirow{2}{*}{29.05} \\
& Where was my bag after I entered the vehicle? & On the passenger seat. & \\
\bottomrule
\end{tabular}
\caption{Utility question categories and examples. The percentages indicate the proportion of collected question--answer pairs assigned to each category.}
\label{tab:utility_annotation}
\end{table*}

\subsection{Formative Observation}
\label{sec:observations}
Data collected in this study revealed recurring patterns in how participants reasoned about visual privacy in their daily lives. We derived these patterns from participants' privacy annotations, the accompanying rationales, and the annotation interviews. Based on these data, we identified four typical observations demonstrated below and discussed their implications for the system design described in Section~\ref{sec:system-design}.

\textbf{O1: privacy preferences are highly personalized and context dependent.}
Participants usually exhibited substantial variation in what they considered privacy sensitive, even regarding the same visual content. Although certain privacy concerns were shared across most participants, such as the CVC code on credit cards, many annotations reflected distinct individual preferences. For example, some participants considered shipping labels on delivered packages to be privacy-sensitive because they revealed personal information such as names and home addresses (P2, P5, P10), whereas others regarded unopened packages as acceptable to share (P1, P8). We also observed that even for the same individual, privacy preferences regarding the same objects were not fixed. Similar objects were often treated differently depending on surrounding context or object status. For instance, computer screens displaying routine desktop content were generally considered acceptable (P7, P9), whereas screens showing emails (P2), SSH sessions(P2), private conversations (P4, P6), or just desktops but featuring personally preferred animated characters (P8) were consistently marked for obfuscation. These observations suggest that privacy cannot be adequately represented by a universal set of sensitive object categories. Instead, effective privacy decisions should also account for both individual user preferences and contextual states.

\textbf{O2: privacy judgments depend on semantic meaning rather than object categories.}
According to the rationales provided by participants, we found that participants rarely judged privacy based solely on the physical object categories themselves. Instead, they focused on the information those objects revealed. For example, computer screens were not considered privacy-sensitive simply because they were screens. Participants consistently regarded them as non-sensitive when they were blank or powered off, but marked them once they displayed work-related content, emails, or private conversations (P2, P4, P6).
Likewise, documents were annotated because they exposed personal information such as names, home addresses, billing details, or signatures, rather than because they were pieces of paper (P1, P2, P6). Wi-Fi routers were considered sensitive only when their labels revealed network credentials (P2), and identification badges were marked because they disclosed personal identity rather than because they were cards or wearable accessories (P4, P10). We also observed that participants sometimes annotated human body parts not because of the body parts themselves, but because of the actions they revealed. For example, one participant wanted to obfuscate their legs while shaking because the participant viewed this unconscious behavior as a bad habit that could negatively affect others’ perceptions of them, even though legs themselves were not regarded as privacy-sensitive (P2). Similarly, another participant chose to obfuscate their hands while cracking their knuckles because they regarded knuckle cracking as a bad habit (P8). These observations suggest that privacy judgments are driven primarily by the semantic meaning and information exposed by visual content, rather than by predefined object categories alone. Consequently, desired privacy obfuscation systems need to understand what information is revealed under the scene, rather than merely detecting which objects are present.

\textbf{O3: participants prefer fine-grained privacy obfuscation while preserving useful context.}
Instead of removing entire frames or whole objects, participants preferred obfuscating only the specific regions that exposed sensitive information while preserving the surrounding scene. For instance, participants often chose to mask only a profile picture instead of the entire Instagram page (P2), or only the label on a pill bottle rather than the bottle itself (P1, P7). We also observed cases where participants selected only the logo on a steering wheel (P5), leaving the rest of the object visible for subsequent analysis or sharing. These examples suggest that privacy-sensitive information is often confined to a small portion of an otherwise non-sensitive object. 
This preference was also reflected in the utility annotations. Participants generally expected downstream tasks to reason about objects, activities, and their relationships rather than the small privacy-sensitive details themselves. They therefore preferred preserving as much visual context as possible while removing only the information they considered sensitive.  Hence, advanced privacy preservation systems should support fine-grained region-level obfuscation instead of masking entire objects or images, thereby protecting privacy while maintaining downstream utility.

\textbf{O4: privacy preferences cannot be fully predefined and fixed in advance.}
Although participants generally expressed consistent privacy preferences throughout the study, we observed that they did not always identify every privacy-sensitive region during the initial pass of annotation. In some cases, participants noticed that they had missed something and returned to earlier annotations to add previously overlooked regions or revise prior decisions. Therefore, privacy preferences are inherently open-ended and cannot always be fully articulated in advance. Instead, privacy preference elicitation is an iterative process in which participants gradually refine their understanding of what they consider privacy-sensitive as they encounter new activities or environments.

\section{System Design}
\label{sec:system-design}

Our formative findings in Section~\ref{sec:observations} highlight the need for personalized privacy preservation that obscures visual information according to individual users’ preferences, while preserving as much of the remaining visual context as possible for downstream analysis. Addressing this trade-off requires privacy decisions to be precisely localized to the relevant visual regions and informed by what those regions reveal in context. Although recent general-purpose VLMs provide strong semantic understanding, they still struggle with fine-grained region-level grounding tasks~\cite{ma2024groma,yang2023set}. We therefore explicitly separate visual localization from privacy reasoning, employing a dedicated localization mechanism to identify candidate visual regions, so that each subsequent privacy decision can be directly associated with the region to which it applies.

However, localization only tells us where a region is, but not what information it reveals in context. Our formative observations (\textbf{O1}, \textbf{O2}) show that privacy decisions depend on the information revealed by a region rather than on object identity alone, and that similar information can lead to different decisions across users. Thus, interpreting what a region reveals and deciding whether it should be obscured are distinct reasoning problems. We separate semantic interpretation from personalized privacy reasoning so that visual evidence can be characterized without conditioning on the current user’s preferences. Semantic interpretation describes what a region reveals in context, while personalized privacy reasoning determines whether that information should be obscured according to the current user’s preferences.

Personalization introduces a further challenge because privacy preferences are difficult to specify exhaustively in advance. Prior work has shown that visual privacy decisions vary across contexts and, more broadly, that one-time privacy settings may not adequately capture such variation~\cite{hoyle2014privacy,wijesekera2018contextualizing}. Consistent with these findings, our formative observation (\textbf{O4}) suggests that users may express additional privacy preferences as they encounter new situations. A personalized privacy system is then supposed to continue acquiring additional user-specific preference evidence over time. However, exhaustively requesting explicit confirmation would introduce repeated user interaction and impose substantial burdens. In this case, we designed a selective privacy preference acquisition mechanism so that incorporates newly confirmed preferences without requiring extensive confirmation, then enabling these preferences to inform subsequent privacy decisions.

Based on these considerations, we define four design goals that guide the design of \systemname.
\begin{itemize}
    \item \textbf{D1:} Privacy-sensitive visual information should be selectively obscured at the region level while preserving as much surrounding visual context as possible.
    \item \textbf{D2:} Privacy decisions should build on what visual content reveals in context rather than object identity alone.
    \item \textbf{D3:} Privacy decisions should align with individual users’ preferences.
    \item \textbf{D4:} Personalization should selectively acquire and incorporate additional privacy preferences without request confirmation for every privacy decision.
\end{itemize}
Guided by these design goals, \systemname organizes personalized privacy preservation into two stages, as illustrated in Figure~\ref{fig:teaser}. The first stage (Figure~\ref{fig:teaser} top) establishes the spatial and semantic basis for subsequent privacy reasoning. A dedicated localization mechanism identifies candidate visual regions that serve as the spatial units for privacy decisions, enabling region-level obfuscation while preserving surrounding visual context (\textbf{D1}). A VLM then interprets what each region reveals in context and produces a preference-agnostic semantic observation (\textbf{D2}). Rather than merely identifying the object or region, the semantic observation describes the information that the region exposes in the surrounding visual context. The VLM receives no information about the current user’s privacy preferences and is not asked to assign any privacy action, thereby keeping semantic interpretation separate from the personalized privacy decision made in the second stage.

The second stage (Figure~\ref{fig:teaser} bottom) performs personalized privacy reasoning over these semantic observations. Prior work on language-model personalization has explored retrieval-based approaches~\cite{salemi2024lamp}, which incorporate information retrieved from a user’s history into prompts, and parameter-based approaches~\cite{tan2024democratizing}, which adapt model parameters using personalization data. Recent work suggests that these mechanisms can provide complementary benefits when combined~\cite{salemi2025comparing}. Inspired by this work, \systemname combines shared task adaptation with explicit retrieval of user-specific preference evidence. Rather than learning separate parameters for individual users, we use parameter-efficient fine-tuning on preference data pooled across users to specialize an open-source LLM for privacy reasoning.
To smartly acquire additional preference evidence without requesting confirmation for every privacy decision, we designed the \textit{Ask} mechanism that selectively requests user input when the model cannot confidently distinguish the privacy decisions between \textit{Mask} and \textit{Keep}. Confirmed decisions are incorporated into the user's preference memory and can influence the subsequent inferences without retraining the model.  The privacy reasoning model and explicit user-specific memory enable \systemname to tailor privacy decisions to each user’s expressed preferences (\textbf{D3}), and the \textit{Ask} mechanism can effectively concentrate user feedback on decisions that are more likely to be incorrect, reducing unnecessary confirmation requests while acquiring additional preference evidence (\textbf{D4}).

\section{System Implementation}
\label{sec:system-implementation}
We implement \systemname as a two-stage pipeline comprising user-independent semantic understanding and personalized privacy reasoning, with wearer feedback supporting preference adaptation.

\subsection{User-Independent Region-Level Semantic Understanding} \label{subsec:sem-understand}
\subsubsection{Semantic Region Segmentation} \label{subsubsec:segmentation}
\begin{figure*}
    \centering
    \includegraphics[width=1\linewidth]{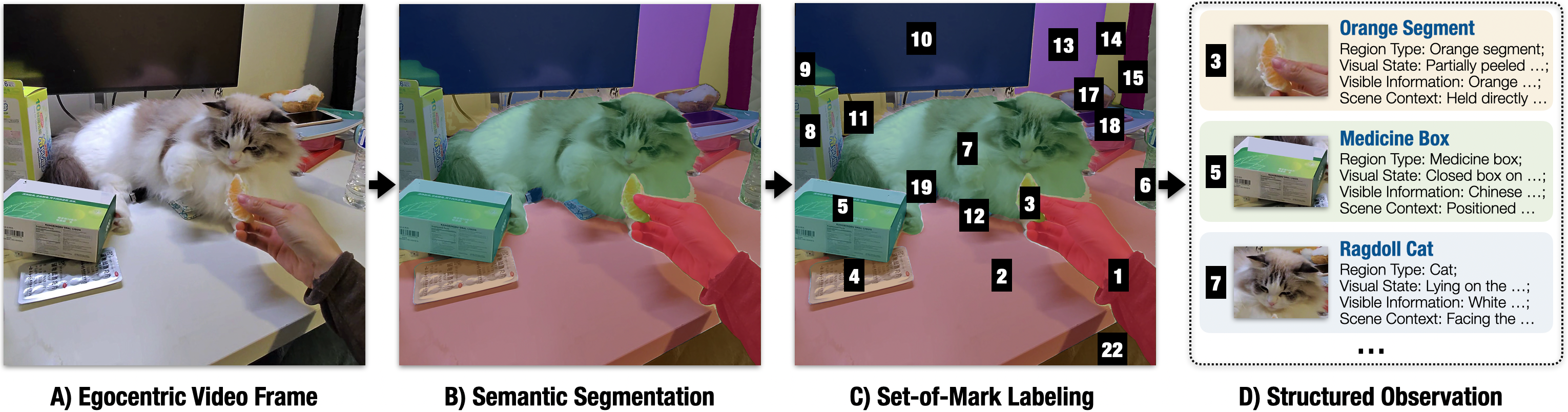}
    \caption{User-independent semantic understanding in \systemname. A) An example egocentric video frame. B) Semantic-SAM segments the frame into semantically coherent visual regions. C) Set-of-Mark labeling assigns numerical identifiers to the visual regions for visual grounding. D) A vision-language model generates structured, user-neutral observations of the visual content in each region.}
    \label{fig:semantic-understanding}
\end{figure*}

For each egocentric frame, we first reuse the segmented regions described in Section~\ref{para:privacy anno} as the semantically coherent but non-overlapping visual regions. 
We assign each of the $N$ resulting regions a unique numerical identifier $i \in \{1, \cdots, N\}$ following the Set-of-Mark (SoM) prompting approach~\cite{yang2023set}. SoM has been proven to facilitate visual grounding in VLMs and has been widely adopted in a range of visual understanding tasks~\cite{huang2025accessibility, li2025ferret, yang2025magma, wang2025embodied}. Figure~\ref{fig:semantic-understanding} illustrates this transformation from an egocentric frame to the visual region representation used for subsequent semantic understanding.

\subsubsection{Structured User-Neutral Representation} \label{subsubsec:structured_representation}
The segmented visual regions obtained from Section~\ref{subsubsec:segmentation} do not explicitly encode the semantic information conveyed by their visual content. We employ open-source vision-language models (VLMs) to extract region-level semantic information from these regions. The VLMs are entirely deployed on local edge devices to reduce the risk of privacy leakage. For each frame, we provide the SoM-overlaid image together with the corresponding raw image to the VLM in a single inference, allowing all numbered regions to be interpreted jointly while avoiding redundant processing. The VLM then generates a structured semantic observation $o_i$ for each numbered region $i$, based on its visible content and surrounding scene (see Figure~\ref{fig:semantic-understanding} for examples). The complete prompt is provided in Appendix~\ref{app:prompt-observation}. The resulting observations follow the schema summarized in Table~\ref{tab:observation-schema}. To ensure that these observations remain independent of wearer-specific privacy preferences, we neither provide the VLM with preference information nor fine-tune it to accommodate such preferences. Instead, the prompt explicitly constrains the VLM to describe only visually observable information, leaving privacy judgments entirely to the subsequent personalized privacy reasoning stage.

\begin{table}[t]
\centering
\small
\begin{tabular}{@{}p{0.22\columnwidth}p{0.38\columnwidth}p{0.28\columnwidth}@{}}
\toprule
\rowcolor{paleblue!90}
\textbf{Field} & \textbf{Description} & \textbf{Example} \\
\midrule
Region Type & Visual entity represented by the region & Shipping Label \\
Visual State & Directly observable state of the region & Partially readable \\
Visible Information & Concrete information visibly conveyed by the region & Recipient name, home address \\
Scene Context & Immediate context surrounding the region & Package near a home entrance \\
\bottomrule
\end{tabular}
\caption{Structured observation schema used for user-neutral semantic representation.}
\label{tab:observation-schema}
\end{table}

\subsection{Personalized Privacy Reasoning} \label{subsec:privacy-reason}

\subsubsection{User Preference Memory and Retrieval}
\label{subsubsec:preference-memory}
For each user $u$, we maintain an explicit preference memory $\mathcal{M}_u$ that stores privacy preferences that the user has previously confirmed. Each memory record $m_{u,j} \in \mathcal{M}_u$ is composed of a structured semantic observation $o_j$ along with its corresponding confirmed privacy action $a_j$ and an optional rationale $\rho_j$ if the action is \textit{Mask}, denoted as 
$$m_{u,j} = (o_j, a_j, \rho_j).$$
Here, $o_j$ has the same structure as the current observation $o_i$ as we obtained from Section~\ref{subsubsec:structured_representation}. We use $i$ to index the current region being evaluated and $j$ to index a historical observation stored in memory.
However, as the preference memory accumulates, many prior preferences may not be pertinent to the current visual context. Providing all stored records would introduce irrelevant information and increase computational overhead. In this case, we use the current observation $o_i$ to retrieve only a small set of the most semantically relevant records from $\mathcal{M}_u$. To effectively retrieve relevant preference memories, we first encode the current observation and each stored memory into a shared semantic embedding space and $\ell_2$-normalize the resulting embeddings, then measure their semantic similarity using cosine similarity
$$
sim(o_i,m_{u,j}) = \frac{e(o_i)^\top e(m_{u,j})}{\|e(o_i)\|_2\|e(m_{u,j})\|_2},
$$
where $e(\cdot)$ denotes the semantic embedding encoder and $sim(o_i,m_{u,j})$ measures the semantic similarity between the current observation $o_i$ and the historical observation $o_j$. We retain only records whose similarity exceeds a minimum threshold $\tau$ and retrieve up to the $k$ highest-scoring records,
$$
\mathcal{R}_u(o_i) = \operatorname{TopK}_{m \in \mathcal{M}_u,\; s(o_i,m)\geq\tau} s(o_i,m),
$$
where $\mathcal{R}_u(o_i)$ denotes the resulting set of retrieved preference records and $|\mathcal{R}_u(o_i)| \leq k$.

In our implementation, we use a frozen \textit{E5-base-v2}~\cite{wang2022text} sentence encoder and retrieve up to $k=3$ preference records for each observation. We set the cosine-similarity threshold to $\tau=0.7$ and records with similarity below this threshold are excluded from retrieval. Consequently, $\mathcal{R}_u(o_i)$ may be empty when none of the wearer's prior preference records are sufficiently related to the current observation. Embeddings of historical observations are computed when records are added to the memory index and reused for subsequent queries, while only the current observation is encoded during inference. This is to avoid repeatedly encoding the entire memory bank for each query and reduces retrieval-time computation as the memory grows.

The retrieved set $\mathcal{R}_u(o_i)$ provides user-specific evidence about how semantically related information was previously treated. Each retrieved record contributes its historical observation, confirmed privacy action, and rationale when available. These records do not directly determine the privacy action for the current region; instead, they are passed together with the current observation to the privacy reasoning model that will be discussed later.

\subsubsection{Privacy Preference Reward Modeling} \label{subsubsec:reward-model}
Retrieved preference records provide concrete evidence about how the current user treated semantically related information in the past, but they do not directly specify how those preferences should apply to the current observation. The current context may differ from the retrieved examples, and relevant user-specific evidence may also be limited or unavailable. We then use a privacy reasoning model to interpret the current semantic observation together with any retrieved preference records and evaluate the candidate privacy actions. 

We formulate this reasoning process as a pairwise reward modeling problem, following the standard scalar preference-reward formulation used to learn from preferred and rejected alternatives. For each annotated observation, we treat the privacy action selected by the user as the preferred action and the other action as the rejected alternative, and train the reward model to assign a higher score to the preferred action. 
For each region and its corresponding observation $o_i$, the privacy action selected by the wearer could be either \textit{Mask} or \textit{Keep}. Here we denote the selected action as 
$a_i^+$, while the alternative action is automatically marked as the rejected action $a_i^-$. 
Following the Bradley-Terry model for pairwise comparisons~\cite{bradley1952rank}, we optimize the difference between their scalar reward scores. 
Given a training set $\mathcal{D}=\{(o_i,a_i^+,a_i^-)\}_{i=1}^{M}$ containing $M$ pairwise preference examples, the objective function is
$$
\mathcal{L}_{\mathrm{RM}} = -\frac{1}{M} \sum_{i=1}^{M} \log \sigma\left(r_\theta(o_i,a_i^+) - r_\theta(o_i,a_i^-)\right).
$$
where $\sigma(\cdot)$ denotes the sigmoid function.

We implement the reward model using Qwen3-8B as the decoder-only language backbone and add a trainable linear reward head that maps a textual input to a scalar reward rather than generating text. Fully fine-tuning the pretrained language-model backbone is computationally and memory intensive. Therefore, we use Low-Rank Adaptation (LoRA)~\cite{hu2021lora} to specialize the model for privacy reasoning while keeping the pretrained model parameters frozen. 
We apply LoRA to the query, key, value, and output projection layers of each self-attention module, with the rank $r=8$, scaling factor $\alpha=16$, and dropout $0.05$. The LoRA parameters and scalar reward head are jointly optimized using the objective function described above, while the rest of the original language-model parameters remain unchanged.

At inference time, the current observation is augmented with user-specific preference records retrieved from $\mathcal{M}_u$. For a candidate action $a$, the model is then fed with $(o_i, \mathcal{R}_u(o_i), a)$ and produces a scalar score. The resulting scores for \textit{Mask} and \textit{Keep} are used in the next stage to determine the region-level privacy decision.

\subsubsection{Region-Level Privacy Decision}
\label{subsubsec:privacy-decision}

We now have personalized reward scores for the candidate privacy actions. The remaining question is how these estimates should be translated into a
final decision for each visual region. 
Suppose that $a_i^*$ is the action receiving the higher reward for observation $o_i$, denoted as
$$
a_i^* = \underset{a \in \mathcal{A}}{\arg\max}\;
r_\theta\left(o_i,\mathcal{R}_u(o_i),a\right),
$$
where $\mathcal{A}=\{\textit{Mask},\textit{Keep}\}$. 
 
We then measure how strongly the model distinguishes the selected action from the alternative $\bar{a}_i$ using the reward difference as follows:

$$
\Delta r_i = r_\theta\left(o_i,\mathcal{R}_u(o_i),a_i^*\right) -r_\theta\left(o_i,\mathcal{R}_u(o_i),\bar{a}_i\right).
$$
A larger $\Delta r_i$ indicates a stronger separation between the two actions, whereas a small difference indicates that the model assigns them similar scores. However, automatically applying the higher-scoring action would force a decision even when the model exhibits only a weak preference between \textit{Mask} and \textit{Keep}. In such cases, we instead involve the user to resolve the ambiguity before applying the privacy action as well as obtaining more confirmed preferences. Here we introduce an empirical decision threshold $\delta$ and defer cases with insufficient reward separation to the user for review. Then the final region-level privacy action is determined as
\begin{equation*}
  \hat{a}_i = 
  \begin{cases} a_i^*, & \Delta r_i > \delta,\\\textit{Ask}, & \Delta r_i \leq \delta.
  \end{cases}  
\end{equation*}

For regions that satisfy the decision threshold, the higher-ranked action $a_i^*$ is applied directly. Regions that are assigned \textit{Mask} are obfuscated, and \textit{Keep} remains unchanged. When the reward difference does not exceed the threshold, the system assigns \textit{Ask} and presents the region to the user for review before applying a privacy action. The confirmed or corrected preference can then be incorporated into the user's preference memory, as described in the following section.

\subsection{User Feedback and Preference Memory Updates}
\label{subsubsec:user-feedback}

\subsubsection{Feedback Interface}
\begin{figure*}[t]
    \centering
    \includegraphics[width=\linewidth]{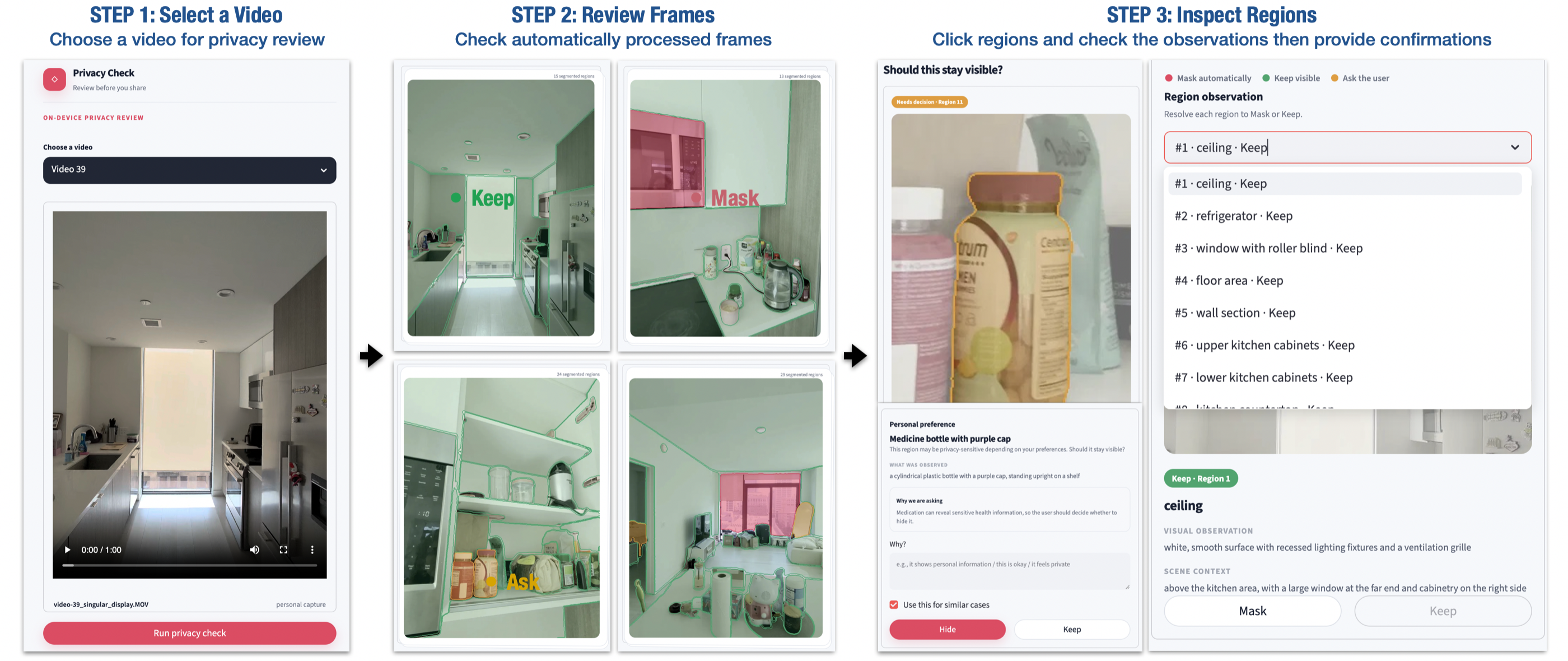}
    \caption{Feedback interface for reviewing privacy decisions on the user's device. Regions assigned \textit{Ask} are proactively presented for system-initiated confirmation, while automatically assigned \textit{Mask} and \textit{Keep} decisions remain available for user-initiated correction. Users can also provide rationales for regions they choose to \textit{Mask}.}
    \label{fig:review_app}
\end{figure*}

System-generated privacy decisions may not always align with users’ current preferences. The system may encounter situations not represented in a user’s existing preference history, and users may also revise preferences they previously expressed. Such mismatches may expose privacy-sensitive information or, conversely, unnecessarily obscure content that users would prefer to retain. We therefore designed a feedback interface that allows users to review privacy decisions on their own device and adjust them as needed (Figure~\ref{fig:review_app}). With this interface, users can restore regions that were unnecessarily obscured, identify missed regions that should be obscured, and provide a rationale for regions they choose to \textit{Mask}.

The interface supports both \textit{system-initiated confirmation} and \textit{user-initiated review}. When the system cannot make a sufficiently confident \textit{Mask} or \textit{Keep} decision, it will assign \textit{Ask} and proactively present the corresponding regions to the user for confirmation as described in Section~\ref{subsubsec:privacy-decision}. 
Automatically assigned \textit{Mask} and \textit{Keep} decisions remain available for user-initiated review, allowing users to either confirm the system’s decision or correct it when it does not align with their preferences. 
The empirical decision threshold $\delta$ determines when \systemname proactively requests user input. Any decision explicitly confirmed or corrected by the user is treated as confirmed preference evidence then incorporated into the preference memory. 

\subsubsection{Preference Memory Maintenance} \label{subsubsec:memory maintenance}
Preferences explicitly confirmed or corrected by the user form the preference memory. However, simply appending these confirmed decisions may introduce redundant or conflicting records that subsequently compete during retrieval. Prior work indicates that preference memories benefit from explicit maintenance to resolve such redundant or conflicting entries~\cite{kirmayr2025carmem}, as semantic similarity alone may be insufficient in distinguishing repetitions or contextual differences~\cite{du2026optimizing}. Inspired by these findings, we apply a lightweight maintenance procedure before incorporating newly confirmed preferences into $\mathcal{M}_u$.

For each confirmed preference, we reuse the semantic embedding and similarity mechanism described in Section~\ref{subsubsec:preference-memory} to retrieve the top three most similar active records from the user’s existing memory. Note that the embeddings are encoded solely from the observations, and the confirmed privacy decisions and their rationales do not contribute to the retrieval similarity. We then use \textit{Qwen3-8B} with a constrained relation-classification prompt to compare the new preference against the retrieved records, and classify the new preference as \textit{Repeated}, \textit{Contextually Distinct}, \textit{Conflicting}, or \textit{Unrelated} with respect to the retrieved preferences. 
The classification results determine how the memory is updated. A \textit{Repeated} preference replaces the corresponding existing one, while \textit{Contextually Distinct} and \textit{Unrelated} preferences are added as new active preferences. If labeled with \textit{Conflict}, the new preference is stored as an active record, while the conflicting one is also retained but marked inactive and excluded from subsequent retrieval. The prompt used for this classification is provided in Appendix~\ref{app:prompt-memory-maintenance}.

\subsection{Inference over Continuous Video}
\label{subsec:inference}
The preceding components describe how privacy actions are determined for individual visual regions within a single frame. Applying semantic understanding and personalized privacy reasoning independently to every region in every video frame would introduce substantial redundant computation, because adjacent frames in egocentric video often contain the same regions with little or no change. Our goal is to avoid repeated reasoning over temporally redundant content while remaining responsive to changes that may affect privacy decisions. To achieve this, we extend the region-level pipeline to continuous video by combining frame-wise region reasoning with temporal tracking using SAM~2.1 with mask prompts. A region that cannot be matched to an active track is treated as newly observed and passes through semantic understanding and personalized privacy reasoning to obtain a privacy action, which is then associated with a new track. A region that remains matched to an active track reuses its current privacy action instead of repeating the full reasoning process in every frame.

\begin{figure*}[t]
    \centering
    \includegraphics[width=\linewidth]{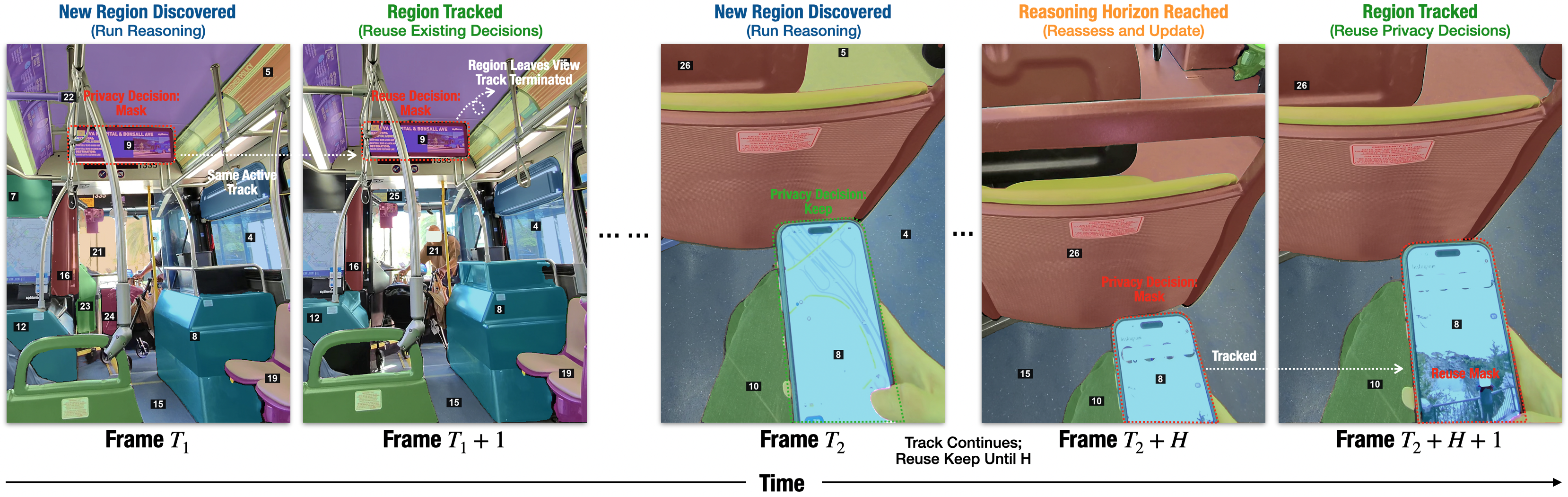}
    \caption{Continuous-video inference in \systemname. Privacy decisions are reused while a region remains actively tracked and are recomputed when the privacy reasoning horizon $H$ is reached. The example illustrates how the same tracked phone screen can change from \textit{Keep} to \textit{Mask} as its displayed content changes from a map to an Instagram interface.}
    \label{fig:continuous-inference}
\end{figure*}

The visual content of a tracked region can nevertheless change in ways that affect its privacy decision. Tracking establishes temporal continuity of the region but does not guarantee that the information it reveals remains unchanged. For example, a computer screen may transition from a blank display to an email while remaining the same physical object. To prevent a privacy action from being reused after the region's semantic content has changed, we introduce a privacy reasoning horizon $H$. A successfully tracked region can reuse its current privacy action for at most $H$ seconds. In our implementation, we set $H$ to 2 seconds. When this horizon is reached, the system regenerates the region's semantic observation and reruns personalized privacy reasoning while preserving the temporal track. If a tracked region disappears, its active track is terminated. A region that later appears without an active track is treated as newly observed and evaluated again. As illustrated in Figure~\ref{fig:continuous-inference}, this procedure reduces repeated reasoning over temporally redundant content while periodically reassessing tracked regions whose privacy-relevant information may have changed.

\section{Evaluation} \label{sec:evaluation}
We evaluate \systemname using an offline simulation that follows the chronological order in which a new user would use the system. Our evaluation examines personalized privacy reasoning on annotated frames, and further studies simulated user feedback and evaluates the privacy-utility trade-off.

\subsection{Experimental Setup} \label{subsec:setup}
\subsubsection{Cross-User Chronological Simulation}
\label{subsec:chronological-simulation}
We conduct a $10$-fold cross-user evaluation in which each user is treated once as a new user of \systemname, while data from the other nine users are used to train the privacy reasoning model. For each held-out user, we simulate real-world system use by processing their video clips in the chronological order in which they were collected. The user begins with an empty preference memory, and privacy decisions for each clip can use only user-specific preference information obtained from earlier clips. All predictions for the current clip are recorded before any annotations from that clip are incorporated into memory, ensuring that current or future preference information cannot influence earlier decisions. Note that all subsequent evaluations follow this cross-user chronological protocol unless otherwise specified.

\subsubsection{Privacy Preference Dataset} \label{subsec:privacy-preference-dataset}
Given this cross-user split, we use the privacy annotations collected during the user study (Section~\ref{sec:user study}) differently for training the privacy reasoning model and for evaluating system performance on the held-out user’s data. 

For the $9$ training users in each fold, each user-selected sensitive region is labeled as a \textit{Mask} training instance. For frames marked as \textit{No Concerns}, we assign the \textit{Keep} label to all segmented regions, as participants confirmed that no visible information in these frames required obscuring. We exclude all \textit{Skip} frames because the corresponding privacy-sensitive content had already been annotated earlier in the same video clip. Since similar non-sensitive content may recur across multiple \textit{No Concerns} frames, the resulting training dataset will contain many semantically redundant \textit{Keep} instances. To reduce this redundancy, we deduplicate only highly similar \textit{Keep} instances within each user’s individual video clips. Specifically, we encode the semantic observation corresponding to each region using Sentence-BERT~\cite{reimers2019sentence} with the \texttt{all-MiniLM-L6-v2} model. We then process the instances in chronological order and remove an instance if its cosine similarity to a previously retained instance exceeds a threshold of $0.7$. Note that all \textit{Mask} instances are retained for the training dataset.

For the remaining $1$ held-out user, we do not perform semantic deduplication. Instead, we run inference on all segmented regions in the annotated frames, excluding those marked as \textit{Skip}. Privacy metrics are then computed based on the region-level labels. For frames containing user-specified privacy-sensitive regions, the selected regions are labeled as \textit{Mask}, while all remaining regions are labeled as \textit{Keep}, as our preprocessing produces disjoint segmented regions. For frames marked as \textit{No Concerns}, all segmented regions are evaluated as \textit{Keep}. Thus, unlike the training data, no redundancy removal is applied to the held-out user’s data, and all eligible segmented regions are included in inference and evaluation.

\subsubsection{Evaluation Metrics}
\label{subsec:evaluation-metrics}
We treat \textit{Mask} as the positive class in privacy reasoning experiments. A true positive (TP) is a privacy decision instance that both the user and the system label as \textit{Mask}, while a false positive (FP) occurs when the system predicts \textit{Mask} for an instance labeled \textit{Keep} by the user. A false negative (FN) denotes the case when the system predicts \textit{Keep} for an instance labeled \textit{Mask}, and a true negative (TN) is an instance labeled \textit{Keep} by both the user and the system. Then the mask precision, recall, and F1 are defined as
\begin{gather*}
\text{Precision} = \frac{TP}{TP+FP}, \qquad
\text{Recall} = \frac{TP}{TP+FN}, \\
\text{F1} = \frac{2 \times \text{Precision} \times \text{Recall}}
{\text{Precision}+\text{Recall}}.
\end{gather*}
Here, precision captures how often the system avoids unnecessary obfuscation, whereas recall measures how often it correctly identifies information that the user wants obscured. Because \textit{Keep} instances are more frequent than \textit{Mask} instances in our dataset, we additionally report balanced accuracy defined as
\begin{equation*}
\text{Balanced Accuracy} = \frac{1}{2}\left(\frac{TP}{TP+FN}+\frac{TN}{TN+FP}\right),
\end{equation*}
which gives equal weight to performance on \textit{Mask} and \textit{Keep} decisions.

We also measured downstream utility using question-answering (QA) accuracy in the full-video evaluation. Each video clip is processed under the same chronological system state used in the privacy evaluation, with access only to user-specific preference information obtained from earlier clips. We apply the same downstream model and prompting procedure to each processed video and measure whether the information required to answer the user’s questions remains available after obfuscation. We do not evaluate region-level privacy decisions on the processed full videos because privacy annotations are available only for the annotated frames. Note that all metrics described above are computed separately for each evaluation user and then macro-averaged across the 10 cross-user folds.

\subsection{Comparative Evaluation of Personalized Privacy Reasoning on Annotated Frames}
\label{subsec:privacy-reasoning}
\subsubsection{Comparison Experiments and Ablation Setup}
We first evaluate two general VLM baselines using Qwen3-VL-8B-Instruct~\cite{bai2025qwen3}. \textit{Generic VLM} receives a Set-of-Mark (SoM) image containing the number IDs and is prompted to directly identify which regions should be obscured. \textit{Generic VLM + Memory} additionally receives preference records retrieved from the evaluation user's chronologically available preference memory. Because \systemname uses the same VLM to generate preference-agnostic semantic observations, these baselines provide a comparison between direct region-level privacy reasoning and \systemname's separation of semantic interpretation from subsequent personalized privacy reasoning. The prompts used for these baselines are provided in Appendix~\ref{app:prompt-generic-vlm}.

We then conduct controlled ablations of the privacy-reasoning stage. All four controlled conditions receive the same preference-agnostic semantic observations and use Qwen3-8B~\cite{bai2025qwen3} as the privacy-reasoning backbone with a single scalar scoring head. \textit{Base Privacy Reasoner} freezes the pretrained Qwen3-8B model without preference memory. \textit{Base Privacy Reasoner + Memory} additionally receives preference records retrieved from the evaluation user's memory. \textit{Shared LoRA Reasoner} uses the same Qwen3-8B backbone with a LoRA adapter trained on privacy preference data from the other nine users in the corresponding cross-user fold for $8$ epochs on 4 RTX A5500 GPUs, but receives no user-specific preference memory. The full system combines the shared LoRA reasoner with retrieved preference records from the evaluation user's memory. All memory-enabled conditions use the same chronologically available preference history and retrieval procedure.

\subsubsection{Privacy Reasoning Performance}
Table~\ref{tab:privacy-reasoning-performance} summarizes the personalized privacy reasoning performance under the cross-user chronological evaluation. \systemname achieves the strongest performance among comparison conditions in all evaluation metrics, with a balanced accuracy of 91.93\%, Mask Precision of 58.01\%, Mask Recall of 84.42\%, and Mask F1 of 63.02\%. These gains matter because an effective obfuscation system must protect sensitive information while still preserving enough context for the video to remain useful. Higher Mask Recall means that \systemname is less likely to leave user-defined sensitive regions exposed. Meanwhile, the gains in balanced accuracy and precision indicate that this privacy protection is not achieved by indiscriminately masking visual content. By better preserving non-sensitive regions, \systemname can reduce unnecessary visual degradation, maintain downstream usefulness, and make the privacy-preserved video easier for users and AI assistants to interpret. Overall, the results suggest that personalized privacy reasoning can improve both privacy protection and the day-to-day usability of egocentric video obfuscation.

\begin{table*}[t]
\centering
\begin{tabular}{lcccc}
\toprule
\rowcolor{paleblue!90}
\textbf{Method}
& \textbf{Bal. Accuracy (\%)}
& \textbf{Precision (\%)}
& \textbf{Recall (\%)}
& \textbf{F1 (\%)} \\
\midrule
\multicolumn{5}{l}{\textit{General VLM baselines}} \\
Generic VLM
& 53.85 & 43.02 & 7.93 & 7.29 \\
Generic VLM + Memory
& 62.10 & 17.60 & 25.35 & 19.36 \\
\midrule
\multicolumn{5}{l}{\textit{Controlled reasoning ablations}} \\
Base Privacy Reasoner
& 74.59 & 15.65 & 60.17 & 19.24 \\
Base Privacy Reasoner + Memory
& 85.23 & 24.34 & 73.99 & 31.37 \\
Shared LoRA Reasoner
& 76.17 & 16.01 & 61.56 & 21.33 \\
\textbf{\systemname}
& \textbf{91.93} & \textbf{58.01} & \textbf{84.42} & \textbf{63.02} \\
\bottomrule
\end{tabular}
\caption{Personalized privacy reasoning performance on annotated frames under the cross-user chronological simulation. Results are macro-averaged across the 10 evaluation users.}
\label{tab:privacy-reasoning-performance}
\end{table*}

\subsubsection{Ablation Analysis}

The general VLM baselines show the limitation of directly making fine-grained privacy decisions from visual input. Generic VLM achieves only 7.93\% recall when directly predicting privacy-sensitive regions from the SoM image. This setting requires the VLM to jointly ground the numbered regions, interpret the information they reveal, and make privacy judgments in a single inference. In comparison, the Base Privacy Reasoner, which operates on explicit region-level semantic observations, achieves substantially higher recall (60.17\%) and balanced accuracy (74.59\%) even without LoRA adaptation or user-specific memory. This result suggests that separating visual interpretation from privacy reasoning provides a more effective representation for fine-grained privacy decisions. Adding preference memory to the Generic VLM improves recall to 25.35\%, but its overall performance remains substantially below the reasoning-based approaches.

The controlled ablations further demonstrate the importance of user-specific preference memory. Adding memory to the Base Privacy Reasoner improves balanced accuracy from 74.59\% to 85.23\% and recall from 60.17\% to 73.99\%. In contrast, the Shared LoRA Reasoner provides only modest improvements over the Base Privacy Reasoner when used without user-specific memory. Since the Shared LoRA is trained only on preference data from the other nine users, this suggests that cross-user preference training can learn transferable privacy reasoning patterns, but cannot by itself capture the preferences of an unseen user.

Combining the learned reasoner with user-specific memory provides the strongest performance. Compared with Base Privacy Reasoner + Memory, \systemname improves balanced accuracy from 85.23\% to 91.93\%, precision from 24.34\% to 58.01\%, recall from 73.99\% to 84.42\%, and F1 from 31.37\% to 63.02\%. Since both conditions use the same user-specific preference history and retrieval procedure, these gains indicate that the learned reasoner more effectively interprets and applies retrieved preferences to new observations. Finally, these comparison results show that cross-user learned reasoning and explicit user-specific memory provide complementary benefits.

\subsubsection{Performance with Increasing User Preference History} \label{subsubsec:performance_history}
We further examine how privacy reasoning performance changes as more user-specific preference history becomes available. Simply comparing earlier and later clips would make it unclear whether performance differences come from having more preference history or from differences in the privacy-sensitive content across clips. To isolate the effect of preference history, we therefore evaluate the same set of clips while varying only how much preference history from earlier clips is available.

For each evaluation user, we divide the chronologically ordered video clips into five contiguous groups of approximately equal size. The final group serves as a fixed evaluation set, while the preceding four groups provide progressively more preference history. We evaluate the final group under five conditions: with no user-specific preference memory, and with preference records accumulated from the first one, two, three, or all four preceding clip groups. Thus, every condition is evaluated on the same clips, while only the amount of prior user preference history changes. Across all conditions, we keep the privacy reasoning model, semantic observations, and retrieval procedure fixed. Preference information from the final evaluation group is not incorporated into memory in any condition.

\begin{figure}[t]
    \centering
    \includegraphics[width=1\linewidth]{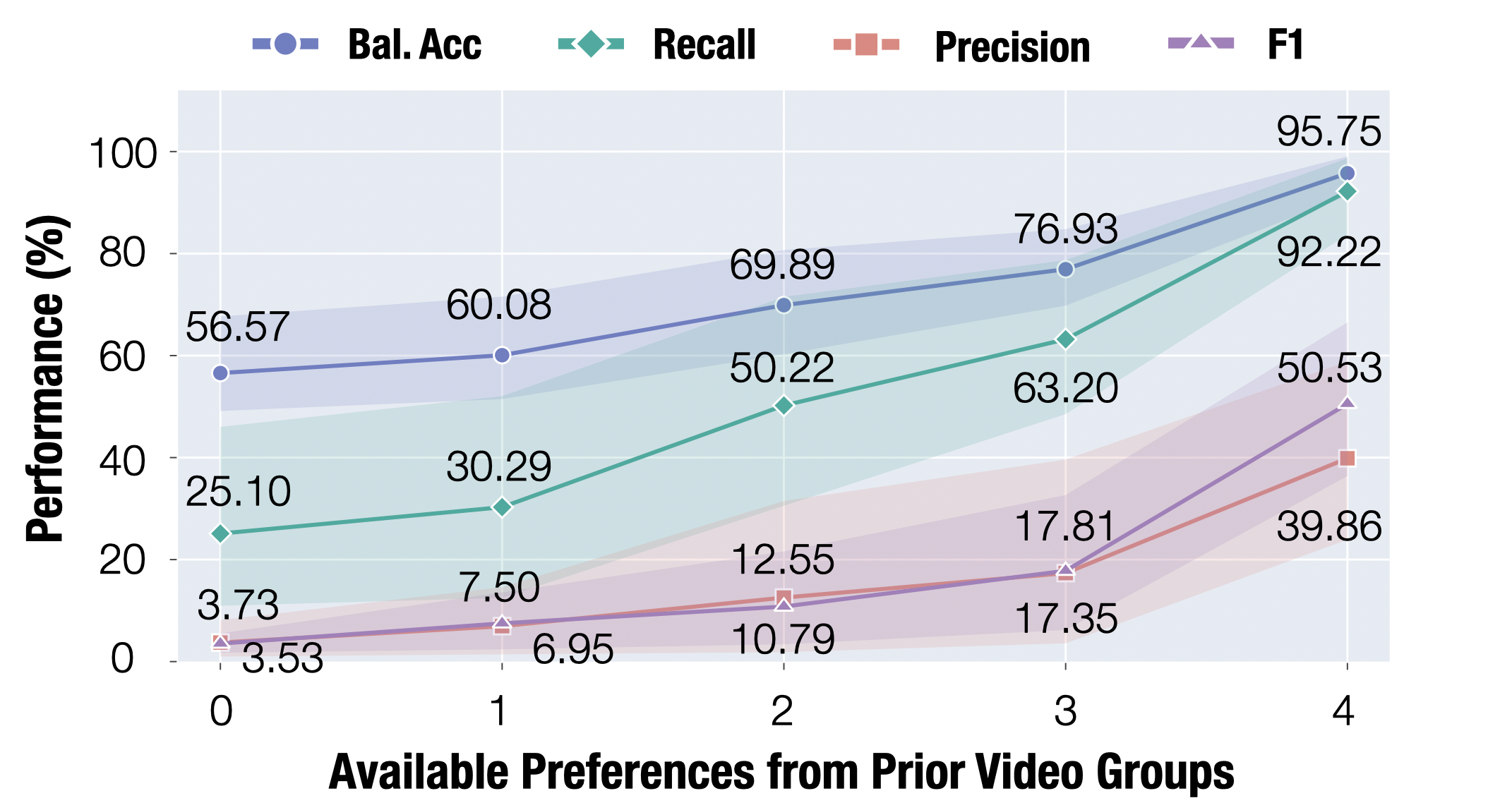}
    \caption{Privacy reasoning performance with increasing amounts of user-specific preference history. For each evaluation user, chronologically ordered clips are divided into five contiguous groups. The final group is held fixed for evaluation, while preference memory is constructed from progressively more of the preceding four groups. Curves show the macro-average across the 10 evaluation users, and shaded regions indicate 95\% user-level bootstrap confidence intervals.}
    \label{fig:increasing-history}
\end{figure}

Figure~\ref{fig:increasing-history} shows Balanced Accuracy, Precision, Recall, and F1 as the amount of available preference history increases. Each curve reports the macro-average across the 10 evaluation users, with shaded regions indicating 95\% user-level bootstrap confidence intervals. Performance consistently improves as more preference history becomes available. From no preference memory to the full four-group history, balanced accuracy increases from 56.57\% to 95.75\%, precision from 3.73\% to 39.86\%, recall from 25.10\% to 92.22\%, and F1 from 3.53\% to 50.53\%. The largest gains across all four metrics occur when the full preceding history is available. These results show that accumulating wearer-specific preference evidence substantially improves privacy reasoning on the same held-out content.

\subsection{Feedback Simulation Using User Annotations}
\label{subsec:simulated-feedback}
The preceding evaluations assume that all user preferences expressed in earlier clips have been confirmed and are available to the system. In practice, however, requiring users to confirm every privacy decision would result in repeated interaction demands and increase user burdens. In this section, we examine the effectiveness of the \textit{Ask} mechanism and evaluate whether the acquired preferences improve privacy reasoning on subsequent clips. Since our study is conducted entirely offline, we simulate user responses using the corresponding privacy annotations and quantify feedback demand by the number of requests. For each clip, all predictions are recorded before the confirmed preferences from that clip are incorporated into memory, ensuring that feedback can affect only subsequent clips.

\subsubsection{Effectiveness of the Reward-Margin-Guided \textit{Ask} Mechanism}
\label{subsubsec:ask-effectiveness}

The \textit{Ask} mechanism is designed to selectively acquire user preferences, particularly when the user's preference memory is still limited. Requiring confirmation for every privacy decision would provide more preference evidence but at the cost of frequent user interaction. To balance preference acquisition with user burden, we designed \textit{Ask} using the reward margin to select a subset of decisions for confirmation. These requests should ideally focus on cases where user confirmation is needed, rather than on decisions that the system can already make correctly.

In this experiment, we vary the effective decision threshold $\delta$ to obtain target feedback-request rates of 5\%, 10\%, 20\%, 30\%, and 40\%. A lower request rate corresponds to a more conservative interaction setting, in which the system asks only about the smallest-margin decisions where it is least confident. For example, at a 5\% request rate, only the 5\% of decisions with the smallest absolute reward margins are sent to the user for confirmation. As request rate increases, progressively more decisions with larger reward margins are involved. Varying the request rate lets us examine the trade-off between minimizing user interruptions and acquiring more confirmed preference evidence. For each held-out user and each request rate, we rank all decisions by their absolute reward margin and select the required number with the smallest margins.

At each request level, we compare \textit{Ask} with a matched random-selection baseline that uniformly samples the same number of decisions from the same held-out user. For each selected decision, we compare the system's automatic \textit{Mask} or \textit{Keep} prediction with the corresponding user annotation. We define the \emph{selected-decision error rate} as the proportion of selected decisions for which the automatic prediction is incorrect. A higher error rate means that the selection strategy directs more of its feedback requests toward decisions that would otherwise result in an incorrect privacy action. In contrast, if the automatic prediction already matches the user annotation, confirmation does not change the current decision. In this experiment, we only focus on whether the \textit{Ask} mechanism chooses more error-prone decisions than random selection, so we do not update the preference memory with the selected feedback here. We evaluate the effect of adding selectively acquired preferences to memory in the next section.

\begin{figure*}[t]
    \centering
    \includegraphics[width=0.8\linewidth]{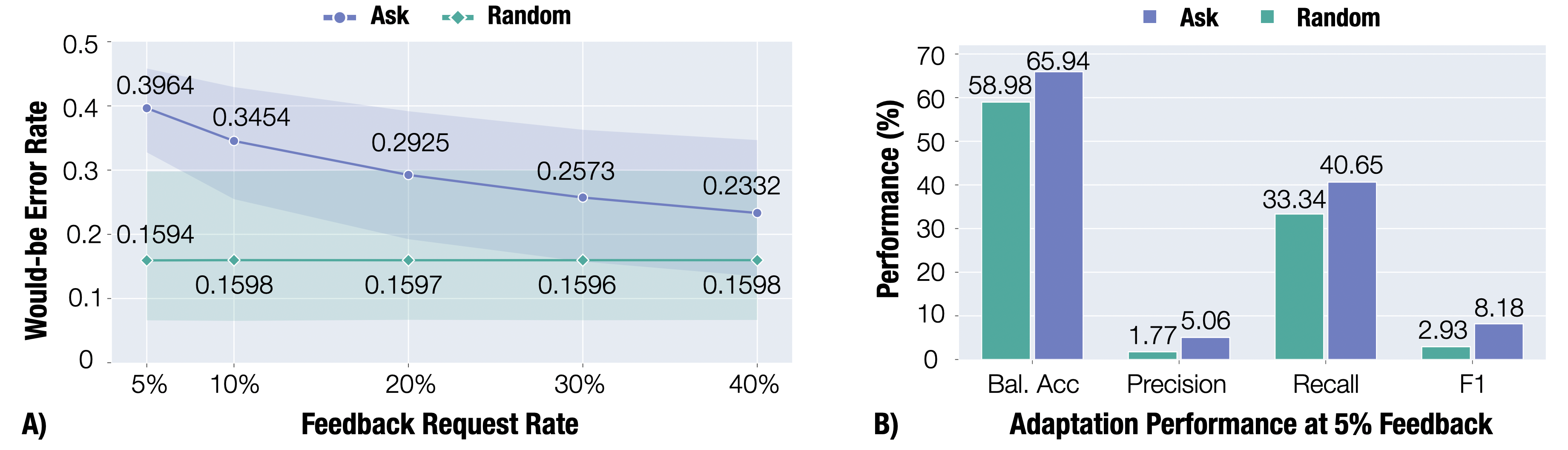}
    \caption{Effectiveness of reward-margin-guided feedback selection and subsequent adaptation. (A) Would-be error rate among decisions selected for feedback across different feedback-request rates. \textit{Ask} selects decisions with the smallest reward margins, while Random selects the same number of decisions uniformly at random. Higher values indicate that feedback requests are more concentrated on decisions the system would otherwise predict incorrectly. (B) Privacy reasoning performance on the fixed G5 evaluation group after acquiring preferences from G1--G4 under a 5\% feedback-request rate. \textit{Ask} and Random receive the same number of user confirmations and differ only in which decisions are selected for feedback.}
    \label{fig:feedback-adaptation}
\end{figure*}

Figure~\ref{fig:feedback-adaptation} (A) shows that the \textit{Ask} mechanism consistently selects more error-prone decisions than random selection across all feedback-request rates. At a 5\% feedback-request rate, 39.64\% of the decisions selected by \textit{Ask} would otherwise be predicted incorrectly, compared with 15.94\% under random selection. In contrast, the random selection remains nearly constant at approximately 16\% across all request rates. 

Results indicate that the model's reward separation is meaningfully aligned with decision correctness, as there are greater would-be errors among the smallest-margin decisions. In other words, decisions the model treats as uncertain are more likely to be ones that would benefit from user correction, while larger-margin decisions are more likely to reflect reliable automatic judgments. This behavior is important because a margin-based mechanism would be less useful if the model were confidently wrong on many decisions or assigned low margins to cases that should be straightforward. 

Furthermore, results verified that the reward margin provides a useful signal for allocating a limited feedback budget, and the declining advantage over random selection is therefore expected. As the request rate increases, \textit{Ask} first exhausts the most ambiguous and error-prone decisions, then progressively includes decisions that the model distinguishes more confidently and that are less likely to be incorrect. Thus, the mechanism provides its strongest error enrichment when user-interaction capacity is most constrained, while also offering evidence that the learned reward margins capture meaningful uncertainty for guiding user feedback.

\subsubsection{Adaptation from Selectively Acquired Preferences} \label{subsubsec:feedback-adaptation}
The last experiment in Section~\ref{subsubsec:ask-effectiveness} examines whether \textit{Ask} can identify decisions that the system would otherwise predict incorrectly. We next examine whether preferences acquired through this selective request is also more useful evidence for subsequent personalization.

For each evaluation user, we use the same five contiguous chronological clip groups introduced in Section~\ref{subsubsec:performance_history}. Each user begins with an empty preference memory and the first $4$ groups are processed sequentially. For each clip, all privacy predictions are made before any feedback from that clip is added. In this experiment, we use the 5\% ask rate as an example condition. The preference confirmations were simulated by the annotations, and only the selected ones through the two selection strategies are available for the privacy decisions of the last group. Figure~\ref{fig:feedback-adaptation} (B) shows that preferences selected via the \textit{Ask} mechanism lead to better subsequent privacy reasoning performance. Even when both conditions receive the same number of confirmations, the preferences selected by \textit{Ask} provide more useful evidence for later decisions.

\subsection{Utility Evaluation on Full Videos}
\label{subsec:video-utility}

In the preceding experiments, we evaluate personalized privacy reasoning on annotated frames. To better understand the impact of our system on real-world applications, we conduct a supplementary study in this section to examine how much information useful for video question answering remained after applying obfuscation to full videos. Unlike the preceding region-level evaluation, we process each video clip using the continuous-video inference procedure described in Section~\ref{subsec:inference} and evaluate the information retained after privacy preservation.
For this full-video evaluation, we assume a fully confirmed chronological preference history, consistent with the setting used in Section~\ref{subsec:privacy-reasoning}.
We compare the utility performance of videos under two conditions. \textit{Raw Video} uses the original, unmodified video, while \textit{Obfuscated Video} applies the personalized privacy decisions produced by the continuous-video inference pipeline. 

We measure video utility using the question-answer pairs collected for each clip during the user study described in Section~\ref{para:utility anno}. We evaluate two video question-answering models, Qwen3-VL-8B-Instruct~\cite{bai2025qwen3} and InternVL3.5-8B~\cite{wang2025internvl3}. For each model, we use the same questions and prompting procedure for the raw and obfuscated videos. Generated answers are manually checked against the corresponding participant-confirmed answers by a researcher. We use semantic correctness rather than requiring an exact lexical match to the reference answer. An answer is considered correct when it captures the core semantic information required by the question, even if finer-grained details such as a brand or model are omitted or less specific. When a particular attribute is explicitly queried, such as a number, count, time, or color, the corresponding attribute must be correct. The prompts used in this evaluation are provided in Appendix~\ref{app:prompt-utility}.

We uniformly sample $5$ videos from each of the $10$ users, resulting in a total of 50 videos with the amount of 150 Question-Answer pairs. Sampling is performed independently within each user so that every user contributes equally to the evaluation. Table~\ref{tab:video-utility} reports question-answering accuracy on the raw and privacy-preserved videos, with mean and standard deviation computed across users. According to the results, neither model answers all questions correctly even when given the raw video. For \textit{Qwen3-VL-8B}, accuracy decreases from 49.33\% on the raw videos to 42.67\% on the obfuscated videos. \textit{InternVL3.5-8B} shows a smaller decrease that from 55.33\% to 52.00\%. Overall, privacy preservation leads to some loss in downstream question-answering performance, but the magnitude of the decrease is limited in this evaluation, suggesting that most of the information they can successfully recover from the original videos remains available after personalized obfuscation.

\begin{table}[t]
    \centering
    \begin{tabular}{lcc}
    \toprule
    \textbf{Video Condition} &
    \textbf{Qwen3-VL-8B (\%)} &
    \textbf{InternVL3.5-8B (\%)} \\
    \midrule
    Raw Video        & 49.33 $\pm$ 24.98 & 55.33 $\pm$ 22.67 \\
    Obfuscated Video & 42.67 $\pm$ 24.38 & 52.00 $\pm$ 19.06 \\
    \bottomrule
    \end{tabular}
    \caption{Question-answering accuracy on raw and privacy-preserved full egocentric videos. Results report mean $\pm$ SD across 10 users, with five videos sampled per user, yielding 50 videos and 150 participant-authored questions in total.}
    \label{tab:video-utility}
\end{table}

\section{Discussion}
\label{sec:limitations}
As with all technologies, our proposed one has limitations and so does the evaluation of our work, which we acknowledge here to inspire ideas and encourage future work.

\textbf{Selective feedback from reward margin.} 
Although the reward model provides only modest gains in overall privacy prediction performance, it also enables selective feedback through the reward margin. Our results show that \textit{Ask} uses this signal to concentrate user confirmations on decisions containing substantially more would-be errors than random selection, especially under small and limited feedback budgets. This allows the system to direct limited user attention towards decisions where confirmation is most needed, reduce unnecessary interaction while preserving opportunities to correct errors.

\textbf{Scale and diversity of user preference data.}
Our study only recruited 10 users and recorded approximately 6.1 hours of egocentric video. While this dataset provides repeated privacy decisions from each user and supports controlled cross-user and chronological evaluation, the limited number of users and recording duration may not capture the broader range of privacy preferences and everyday contexts that could arise in long-term use. In addition, privacy annotations were collected only from selected frames instead of densely throughout each video, so our privacy evaluation is restricted to explicitly annotated decisions and did not measure and report full-video privacy accuracy. Future work needs to involve a larger and more diverse set of participants over longer periods and investigate scalable approaches for collecting privacy preferences in continuous egocentric video.

\textbf{Offline simulation of user feedback.}
The current implementation of the proposed system is computationally intensive and was therefore evaluated offline instead of through real-time interactive use. We also used previously collected privacy annotations to simulate user responses to \textit{Ask}, which enables controlled evaluation of feedback selection and subsequent personalization. However, this setup does not capture perceived interaction burden or users' willingness to respond to ask requests. Supporting real-time interactive use would require additional system engineering, such as an efficient cloud-assisted pipeline or a fully on-device implementation. We leave these deployment optimizations and their evaluation in longitudinal use to future work.

\textbf{Dependence on visual and temporal understanding.}
The system currently relies on accurate region localization and semantic interpretation before making personalized privacy decisions. Privacy-sensitive information that is missed during localization or not captured in the semantic observation may therefore remain undetected by the subsequent privacy reasoner. Continuous-video inference further relies on temporal tracking and a fixed reasoning horizon $H$, so privacy-relevant changes that occur between reassessments may not be detected immediately. Future work could investigate more robust visual grounding and semantic representations, as well as adaptive reassessment triggered by changes in visual content rather than a fixed temporal interval.

\textbf{Computational cost and real-world deployment.}
The current implementation requires multiple stages of visual and language-model inference, resulting in substantial computational cost for continuous video processing. Although temporal tracking reduces redundant reasoning across adjacent frames, the system has not yet been optimized for real-time operation on resource-constrained wearable devices. Future work should explore more efficient models and inference strategies, including model compression and on-device or cloud-assisted deployment, and evaluate their latency and energy requirements in sustained use. To support future research and reproducibility, we will release our project upon publication at \url{https://github.com/hilab-open-source/EgoVeil}.
\section{Conclusion}
\label{sec:conclusion}

We present \systemname, a personalized privacy reasoning system for egocentric video. It separates region localization and preference-independent semantic interpretation from personalized privacy reasoning, combining a shared reasoner trained on preference data pooled across users with an explicit user-specific preference memory. When a decision is uncertain, \systemname selectively requests user feedback and incorporates confirmed preferences into subsequent reasoning without retraining the shared reasoner. The evaluation results show that the shared reasoner and user-specific memory provide complementary benefits, and that selectively acquired feedback yields greater downstream improvement than an equal amount of randomly acquired feedback. Together, these results demonstrate how egocentric visual privacy preservation can move beyond fixed privacy categories by accounting for both contextual visual information and individual user preferences.

\balance
\bibliographystyle{ACM-Reference-Format}
\bibliography{main.bib}

\appendix
\newpage
\section{Prompts}
\label{app:prompts}

\subsection{Semantic Observation Generation}
\label{app:prompt-observation}

\begin{promptsectionbox}
\begin{lstlisting}[style=promptstyle]
You are given two images of the same scene.

Image 1 is the original complete frame. Image 2 is a Set-of-Mark view of the same frame. Colored outlines delineate regions, and each black badge with a white integer names one region. The badge number is a reference ID, not text that belongs to the photographed scene.

Describe every requested region exactly once. Use Image 1 to inspect visual content and Image 2 only to identify each region's extent and ID. Describe the semantic region indicated by the outline, not the badge. Include ordinary objects, surfaces, structural areas, and background regions. Keep each field concise so that no requested region is omitted.

Describe only visually supported facts. If a region is partial, mixed, blurred, occluded, or ambiguous, state that uncertainty instead of guessing. Do not infer preferences, privacy concerns, sensitivity, intent, identity, ownership, or relationships. Do not mention marks, badges, outlines, colors, masks, segmentation, annotation, or the selection process in any description.

Never transcribe literal identifying or sensitive values. Do not quote exact names, addresses, phone numbers, email addresses, usernames, account numbers, license plates, document wording, message text, medication names, QR values, or other readable identifiers. Instead, report semantic information types such as "person name", "address", "message text", "license plate number", or "account information" in visible_information. Use an empty list when a region contains no distinct information-bearing content.

Requested mask IDs: [MASK_IDS]

Return exactly one JSON object and no other text. The regions array must contain exactly one entry for every requested mask ID, no unrequested IDs, no duplicate IDs, and entries sorted by mask_id:

{
  "regions": [
    {
      "mask_id": [MASK_ID],
      "region_type": "short noun phrase naming this region",
      "visual_state": "short description of its visible state",
      "visible_information": ["semantic information type, if any"],
      "scene_context": "short description of its location and relevant surroundings"
    }
  ]
}
\end{lstlisting}
\end{promptsectionbox}

\textbf{Observation Generation Retry.}
We validate the structured output before using it in subsequent stages. If the VLM omits a requested region, returns an unrequested region, or otherwise violates the expected output structure, we repeat the request with the following additional instruction inserted immediately before the requested region identifiers.

\begin{promptsectionbox}
\begin{lstlisting}[style=promptstyle]
A previous pass did not return every requested ID. Return only the IDs requested below and correct this validation problem: [VALIDATION_ERROR]
\end{lstlisting}
\end{promptsectionbox}

\subsection{Preference Memory Maintenance} \label{app:prompt-memory-maintenance}

\begin{promptsectionbox}
\begin{lstlisting}[style=promptstyle]
Classify how one newly confirmed wearer privacy preference relates to the most semantically similar active preferences from the same wearer.

Use exactly one relation:
- Repeated: materially the same visual situation and the same action. The new record should replace one older record
- Contextually Distinct: related visual content, but a meaningful contextual difference makes both records useful.
- Conflicting: materially the same visual situation but different actions. The new record becomes active and one conflicting older record becomes inactive.
- Unrelated: none of the retrieved candidates describes a sufficiently related visual situation.

Do not make a new privacy judgment. Compare only the supplied observations, confirmed actions, and rationales. For Repeated or Conflicting, select exactly one retrieved Memory ID as target_memory_id. For Contextually Distinct, the target may be a retrieved Memory ID or null. For Unrelated, it must be null.

New preference:
[NEW_PREFERENCE]

Retrieved active candidates:
[RETRIEVED_ACTIVE_CANDIDATES]

Return exactly one JSON object and no other text:
{"relation":"one exact relation above","target_memory_id":null}

Replace null with a JSON string containing one retrieved Memory ID only when a target is required. Do not write the word "null" as a string.
\end{lstlisting}
\end{promptsectionbox}

\subsection{Generic VLM Baselines} \label{app:prompt-generic-vlm}
\begin{promptsectionbox}
\begin{lstlisting}[style=promptstyle]
You are given one egocentric Set-of-Mark image. Each black badge with a white integer identifies exactly one candidate visual region. The badge is a reference marker and is not text from the photographed scene.

For every marked region, predict whether the current wearer would want the visible content obscured for privacy (Mask) or left visible (Keep).

Use the following decision procedure:

1. Base the decision on information that is actually visible in the marked region and its visual context. Do not assume hidden or unreadable content.
2. Judge the information revealed by the region rather than the object category alone. An ordinary object is not automatically private, while visible content on or associated with an ordinary object may be private.
3. When relevant confirmed preferences from this wearer are supplied, treat them as user-specific evidence. Apply a preference only when its visual situation is meaningfully similar to the current region.
4. When no relevant preference is available, use general human privacy expectations as a prior. The absence of a memory record is not evidence for either Keep or Mask.
5. Decide each marked region independently. Do not default all regions to the same action merely because personal preference evidence is unavailable.

Choose Mask when the visible content provides reasonable evidence that the wearer would want to control its exposure to another viewer, or when a relevant confirmed preference supports Mask. Choose Keep when the visible content does not provide such evidence, or when a relevant confirmed preference supports Keep.

The following records were retrieved independently for each current mark from strictly earlier video clips. They never describe the current image:

[RETRIEVED_PREFERENCES_BY_MARK]

Requested mark IDs: [MARK_IDS]

Return exactly one JSON object and no other text:
{"mask_ids":[the integer IDs predicted as Mask]}

Include no duplicate or unrequested IDs. An empty array means every requested region is Keep. Do not explain the answer.
\end{lstlisting}
\end{promptsectionbox}

\subsection{Utility Evaluation} \label{app:prompt-utility}
\begin{promptsectionbox}
\begin{lstlisting}[style=promptstyle]
Watch the full video and answer the question using only evidence visible in the video.

Give a concise, direct answer. Do not explain your reasoning. If the video does not provide enough evidence, answer "Not shown."

Question: [QUESTION]
Answer:
\end{lstlisting}
\end{promptsectionbox}

\section{Keyframe Selection Process} \label{app:keyframe}
We developed a motion-compensated, local-change-aware keyframe selection procedure to reduce the annotation burden without setting a fixed global frame budget. This preprocessing was used only for the user study and is not part of the full video privacy reasoning pipeline.

The procedure consists of three stages. First, we identify low-quality frames affected by severe blur, extreme exposure, or probable lens obstruction. Image sharpness is measured using Laplacian variance and Tenengrad~\cite{pertuz2013analysis}, while exposure is assessed using mean intensity and the proportion of clipped pixels. Second, we reduce visual redundancy while accounting for egocentric camera motion. For each usable frame, we match ORB features~\cite{rublee2011orb} with the anchor frame of the current visual run and estimate a partial affine transformation using RANSAC~\cite{fischler1981random}. After alignment, we compare global color differences, local block-level differences, and edge differences. Frames are grouped as near-duplicates only when these differences remain below predefined thresholds and no localized visual change is detected. The highest-quality frame in each run is retained as its representative. Finally, we consolidate short bursts of nearby change candidates and apply an adjacent deduplication pass. Localized changes and boundaries separated by unusable frames are preserved to avoid removing potentially informative content. Algorithm~\ref{alg:keyframe-selection} summarizes the procedure.

\begin{algorithm}[t]
\caption{Motion-Compensated Keyframe Selection}
\label{alg:keyframe-selection}
\begin{algorithmic}[1]
\Require Video frames $\mathcal{F}$
\Ensure Selected keyframes $\mathcal{K}$

\State Compute frame quality and visual descriptors
\State Mark frames with severe blur, extreme exposure, or probable obstruction as unusable
\State Initialize visual runs from the usable frames

\For{each usable frame $F_t$ in chronological order}
    \State Align $F_t$ with the current run anchor using ORB and RANSAC
    \State Compute global, local, and edge differences
    \If{$F_t$ is a near-duplicate and contains no localized change}
        \State Add $F_t$ to the current run and retain its highest-quality representative
    \Else
        \State Start a new visual run at $F_t$
    \EndIf
\EndFor

\State Consolidate temporally adjacent change candidates
\State Merge adjacent near-duplicate candidates while preserving localized changes and unusable-frame boundaries
\State \Return the remaining representatives as $\mathcal{K}$
\end{algorithmic}
\end{algorithm}

\end{document}